\documentstyle[aps,prl,epsfig,multicol,amssymb]{revtex}
\epsfclipon

\begin{document}

\newcommand \be {\begin{equation}}
\newcommand \ee {\end{equation}}
\newcommand \bea {\begin{eqnarray}}
\newcommand \eea {\end{eqnarray}}
\newcommand \nn {\nonumber}
\newcommand \la {\langle}
\newcommand \ra {\rangle}
\newcommand \rat {\rangle_{\tau}}
\newcommand \raw {\rangle_{_W}}

\title{\bf Linear and non linear response in the aging regime of the 1D trap model}
\author{E.M. Bertin, J.-P. Bouchaud}
\address{\it Commissariat \`a l'\'Energie Atomique,
Service de Physique de l'\'Etat Condens\'e\\
91191 Gif-sur-Yvette Cedex, France}

\maketitle

\begin{abstract}
We investigate the behaviour of the response function in the one dimensional trap model using 
scaling arguments that we confirm by numerical simulations. We study the average position of the random walk at time $t_w+t$ given that a small bias $h$ is applied at time $t_w$. Several scaling
regimes  are found, depending on the relative values of $t$, $t_w$ and $h$. Comparison with the diffusive motion in the absence of bias allows us to show that the fluctuation dissipation relation is, in this case, valid even in the aging regime.
\end{abstract}

\begin{multicols}{2}

The one dimensional trap model (hereafter denoted as 1DTM) has been the focus of renewed 
attention, both in the mathematical community \cite{Fontes,BenArous}, and also using a 
physicist approach \cite{Bertin02,Monthusnew}. This model was proposed in the 70's to describe the properties of one dimensional disordered conductors \cite{Scher,PhysRep}. Very recently, the 
direct relevance of this model for the dynamics of DNA `bubbles' under torsion was emphasized 
\cite{Hwa}. Although the ``annealed'' version of the model, in which a new trapping time is randomly choosen at each step, is now well documented \cite{Monthus,Barkai-JCP,Barkai-PRL}, the full analysis of the quenched model considered here remains challenging. It was shown in \cite{Bertin02,Monthusnew} that, besides interesting dynamical localization properties, the 1DTM exhibits different time scalings, depending on which correlation function is considered. To be more specific, the probability of not moving between $t_w$ and
$t_w+t$ -- called $\Pi(t_w+t,t_w)$ in \cite{Bertin02} -- and the probability of occupying the same site at $t_w$ and $t_w+t$ -- called $C(t_w+t,t_w)$ -- exhibit different scalings: $\Pi$ scales as $t/t_w^{\nu}$ with $\nu <1$, whereas $C$ behaves as $t/t_w$. Then a natural question arises in this context: what is the relevant time scale that governs the response function of the particle 
to an external bias? This question is interesting in the context of the physical applications mentioned above, and also in the context of the out-of-equilibrium Fluctuation Dissipation Theorem that was much discussed recently \cite{Ritort}.  

Let us first briefly recall the definition of the 1DTM. Consider a one dimensional lattice, 
and associate to each site $i$ a quenched random variable $E_i>0$, the energy barrier, chosen 
from a distribution $\rho(E)$. One follows the evolution of a particle driven by a thermal 
noise at temperature $T$ on the lattice. The particle has to overcome the energy barrier
$E_i$ in order to leave site $i$ and reach one neighbour. This naturally leads to an escape 
rate $w_i$ of site $i$ given by an Arrhenius law $w_i=\Gamma_0\, e^{-E_i/T}$. Note that the 
transition rates are a
coarse grained description of the underlying Langevin dynamics, which is not explicitely 
described in this model. The quantity $\Gamma_0$ is a microscopic frequency scale, that will 
be set to unity in the following. Once that particle has escaped the trap, it chooses one of 
the two nearest neighbouring sites, with probability $q_-$ for the left one and $q_+=1-q_-$ 
for the right one. Two particular cases have already been studied in details in the 
literature, namely the unbiased case
$q_{+}=1/2 $ \cite{Bertin02,Fontes,BenArous}, and the fully directed case  $q_+=1$ 
\cite{Ledou,PhysRep,Compte}.

In order to compute a response function, one has first to define an external field 
to which the response will be associated. The simplest choice is to consider small 
deviations from the unbiased case, i.e.~to introduce a small bias $h$, independent 
from the site, such that $q_{\pm} = (1 \pm h)/2$. Note that this is equivalent, as 
long as linear response is obeyed, to the introduction of a uniform small force field 
$F$, which transforms the transition rates $q_{\pm}$ into $q_{\pm} = q_{\pm}^0 \, 
\exp(\pm Fa/2kT)$. Here $a$ is the lattice spacing, which is set to unity in the following; 
the correspondence is then $h=F/2kT$. Physically, this force is due to the external 
electric field in the case of conductors \cite{Scher}, or to an 
asymmetry of the DNA composition along the chain in the model considered in \cite{Hwa}.

As far as the response is concerned, the two natural choices are the average position 
of the walk after the bias is applied, or the average probability current. Interestingly, 
a formal relation exists between the former and the latter, namely that the current is the
time derivative of the average position. Switching on a small bias $h$ at time $t_w$ and 
measuring quantities at a subsequent time $t_w+t$, we define for a {\it given} sample of 
the disorder the average position $x_h (t,t_w)$ and the total probability current 
$J_h (t,t_w)$ as:
\bea
x_h (t,t_w) &=& \sum_n n\, P_n(t,t_w)\\
J_h (t,t_w) &=& \sum_n \phi_{n-1,n}(t,t_w)
\eea
with $\phi_{n-1,n} \equiv W_{n-1\to n}^h \, P_{n-1} - W_{n \to n-1}^h \, P_n$ 
being the local current. Taking the derivative of $x_h (t,t_w)$ with respect to $t$, 
one has:
\be
\frac{\partial x_h}{\partial t} = \sum_n n \frac{\partial P_n}{\partial t} = 
\sum_n n [\phi_{n-1,n} - \phi_{n,n+1}]
\ee
Assuming periodic boundary conditions on a lattice of size $L$, 
and letting eventually $L \to \infty$, one can show that the sum reduces to 
$\sum_n \phi_{n-1,n}$, finally leading to:
\be \label{posit-current}
\frac{\partial x_h}{\partial t} (t,t_w) = J_h (t,t_w)
\ee
Note that in the 1DTM, transition rates $W_{n\to n\pm 1}^h$ depend only on 
$n$, so that $J_h(t,t_w)=0$ for $h=0$, leading to a value of 
$x(t,t_w)=x(0,0)$ in the absence of bias, for any given sample.
Averaging over the disorder, it is clear that Eq.~(\ref{posit-current}) 
is also valid for the averaged quantities $\la x \ra_h(t,t_w)$ and 
$\la J \ra_h(t,t_w)$. Therefore in the following we shall focus only 
on the average position after a bias is applied.

In this section, we shall give some simple scaling arguments in order to predict the behaviour of $\la x \ra_h (t,t_w)$ as a function of the three variables $t$, $t_w$ and $h$. Interestingly, non trivial regimes appear due to the fact that the limits $h \to 0$ and $t,t_w \to \infty$ cannot be inverted. Note that the case $t_w=0$ was studied in \cite{PhysRep}, where it was shown that
a non trivial crossover line appears in the plane $h,1/t$. Let us recall briefly these scaling arguments, since this will be useful in the following. It is convenient to introduce the typical number $N$ of steps of the walk after time $t$, and to express both $\la x \ra_h$ and $t$ as a function of $N$. It is clear that $\la x \ra_h \simeq Nh$; now considering the typical number ${\cal N}_s$ of sites visited by the walk, ${\cal N}_s$ can be approximately written as the sum of a drift contribution and of a diffusive one:
\be
{\cal N}_s \sim N h + \sqrt{N}
\ee
Consider first the case $Nh \gg \sqrt{N}$, corresponding to ${\cal N}_s \sim Nh$. 
Given that the trapping times $\tau_i$ are distributed according to 
$p(\tau) = \mu/\tau^{1+\mu}$, the sum of $M$ independent variables 
$\tau_k$ behaves as $\sum_{k=1}^M \tau_k \sim M^{1/\mu}$. Since 
each site is visited of order $N/{\cal N}_s$ times, $t$ can be expressed as:
\be
t \sim \frac{N}{{\cal N}_s}\, \sum_{i=-{\cal N}_s/2}^{{\cal N}_s/2} \tau_i 
\sim N {\cal N}_s^{\frac{1-\mu}{\mu}}
\ee
which can be rewritten as $N \sim t^{\mu} h^{\mu-1}$. Finally, $\la x \ra_h$ reads:
\be
\la x \ra_h \sim h^{\mu} t^{\mu}
\ee
The criterion $Nh \gg \sqrt{N}$ translates into $h^{\mu} t^{\mu} \gg h^{(\mu-1)/2}
t^{\mu/2}$, or equivalently $t \gg t_h \sim h^{-(1+\mu)/(\mu)}$. On the contrary, 
if $Nh \ll \sqrt{N}$ then ${\cal N}_s \sim \sqrt{N}$, and $N$ and $t$ are
related through $N \sim t^{2\mu \nu}$, so that:
\be
\la x \ra_h \sim h\, t^{2\mu \nu}
\ee
with $\nu \equiv 1/(1+\mu)$. Therefore, the response is linear in $h$ but non linear in $t$ 
for $t \ll t_h$, and non linear both in $h$ and $t$ for $t \gg t_h$.

Let us now turn to the response in the aging regime $t_w \gg 1$. Then if $t$ is small enough (to be specified later), the walk will essentially evolve within the space region of size $\xi(t_w)$ 
already visited (the diffusion correlation length). The average trapping time
$\overline{\tau}(t_w)$ within this region is defined by:
\be \label{eqn-taueff}
\overline{\tau}(t_w) \sim \int_1^{t_w^{\nu}} \frac{\mu d\tau}{\tau^{1+\mu}} \sim t_w^{\nu (1-\mu)}
\ee
which is the average value of $\tau$ computed from the distribution $p(\tau)$, 
taking into account the natural cut-off induced by the dynamics $t_w^{\nu}$ \cite{Bertin02}. 
Therefore, as long
as the particle does not escape from the initial region, the average position 
should drift at a constant velocity $h/\overline{\tau}(t_w)$ (the lattice 
spacing $a$ is taken as the unit of length), leading to:
\be
\la x \ra_h \sim h\, t / t_w^{\nu (1-\mu)}
\ee
This short time regime is limited by two conditions: first, $|\xi(t_w+t) - \xi(t_w)| \ll \xi(t_w)$, which implies $t \ll t_w$, and also $|\la x \ra_h| \ll \xi(t_w)$, which requires $t \ll t^*$, where $t^*$ is a new time scale defined by $t^* \equiv t_w^{\nu} / h$. Note however that this time scale is only relevant if $t^* < t_w$. If we are in the opposite limit $t \gg t_w$, then $t_w$  no longer plays any role and one recovers the results found above in the particular case $t_w=0$. So one has to distinguish between several regimes, depending on the relative values of $t$, $t_w$ and $t^*$:

\begin{itemize}

\item $t^* \gg t_w$ (or $h \ll h^* \sim t_w^{-\mu \nu}$).
In this case, three different regimes appear:
\bea \label{x-scal1}
\la x \ra_h &\sim& h\, t / t_w^{\nu (1-\mu)} \qquad t \ll t_w \\ \label{x-scal2}
\la x \ra_h &\sim& h\, t^{2\mu \nu} \qquad \quad t_w \ll t \ll t_h \\
\la x \ra_h &\sim& h^{\mu}\, t^{\mu} \qquad \qquad \, \, t \gg t_h
\eea
where $t_h \sim h^{-(1+\mu)/\mu}$ as above.

\item $t^* \ll t_w$ (or $h \gg h^*$).
The behaviour of $\la x \ra_h$ can be summarized as follows:
\bea
\la x \ra_h &\sim& h\, t / t_w^{\nu (1-\mu)} \qquad t \ll t^* \\
\la x \ra_h &\sim& h^{\mu}\, t^{\mu} \qquad \qquad \; \, t \gg t^*
\eea

\end{itemize}
It is interesting to reformulate the above equations in terms of scaling functions:
\bea \label{x-scal-small}
\la x \ra_h &=& h\, t_w^{2\mu \nu}\, f_1 (t/t_w) \qquad h \ll h^* \\ \label{x-scal-large}
\la x \ra_h &=& t_w^{\mu \nu}\, f_2 (ht/t_w^{\nu}) \qquad \, \, h \gg h^*
\eea
with the following asymptotic behaviour for the functions $f_1(z)$ and $f_2(z)$:
\bea \label{f1-behav}
f_1(z) &\sim z& \quad (z \ll 1), \qquad f_1(z) \sim z^{2\mu \nu} \quad (z \gg 1)\\ \label{f2-behav}
f_2(z) &\sim z& \quad (z \ll 1), \qquad f_2(z) \sim z^{\mu} \qquad (z \gg 1)
\eea
The scaling function $f_1(z)$ only accounts for the time regimes described by Eqs.~(\ref{x-scal1},\ref{x-scal2}), i.e. for times smaller than $t_h$. However, since we are in the small bias case $h \ll t_w^{-\mu \nu}$, $t_h$ is much larger than $t_w$, so that this crossover scale is difficult to evidence numerically.

\begin{figure}
\centerline{
\epsfxsize = 8.5cm
\epsfbox{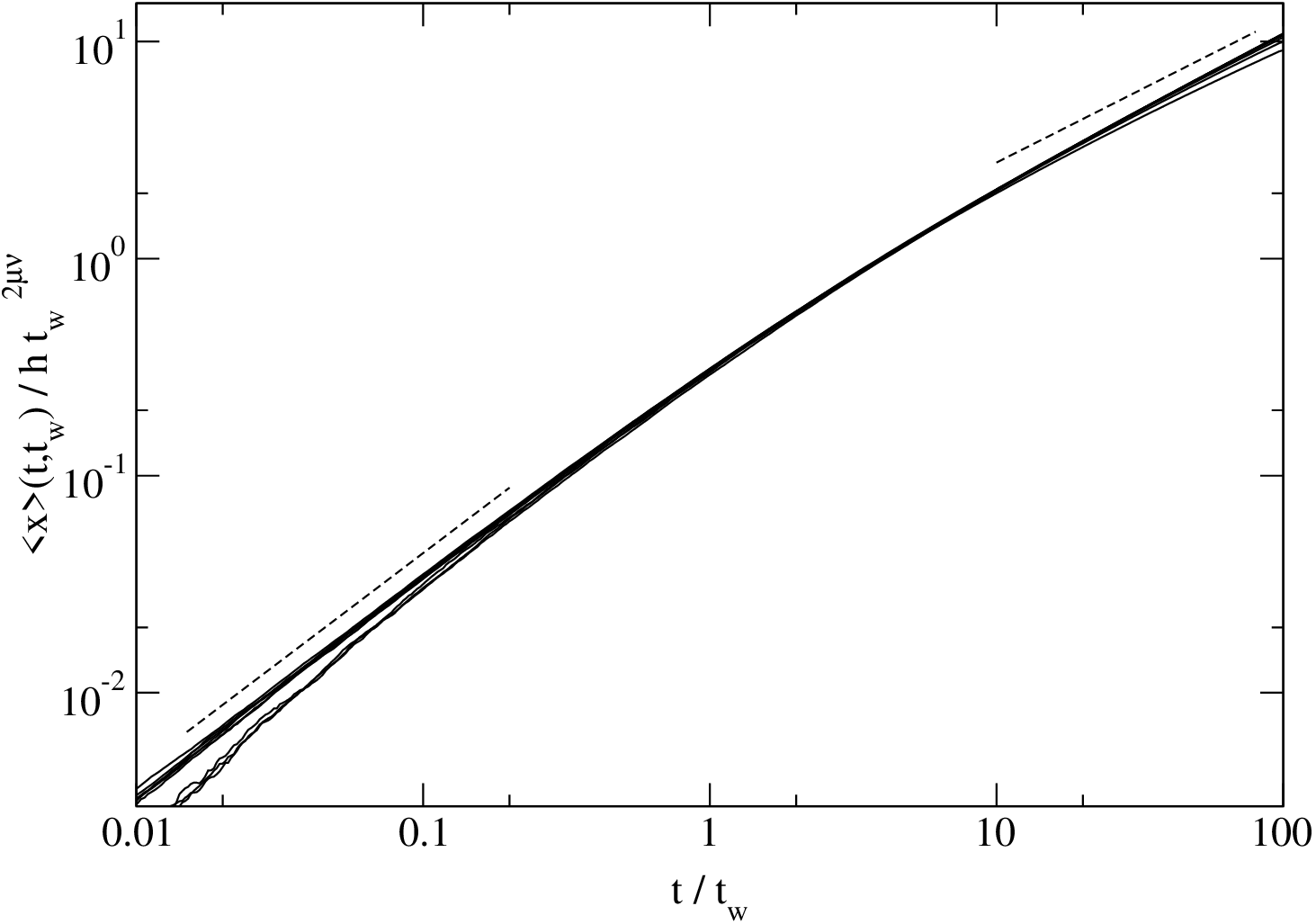}
}
\vskip 0.5 cm
\caption{\sl Rescaled response in the small field regime 
($h \ll h^*(t_w)$), using Eq.~(\ref{x-scal-small}), for 
$h=2.10^{-3}$, $5.10^{-3}$, $8.10^{-3}$ and $t_w=10^3$, $10^4$, 
$10^5$ ($\mu=1/2$). The resulting collapse is good, even though 
some deviations appear both at short and large times. The predicted asymptotic scaling behaviour is well obeyed (dashed lines).
}
\label{fig-small-h}
\end{figure}

We have not found the way to include all the $t$, $t_w$ and $h$ regimes in a single scaling 
function. We now report on the numerical results on the response function, and compare 
them to the 
scaling predictions of the previous section. In the small field regime $h \ll h^*(t_w)$, 
the scaling predicted by Eq.~(\ref{x-scal-small}) is well satisfied. Fig.~\ref{fig-small-h} 
shows the 
resulting collapse of the data for three different values of $h$ ($h=2.10^{-3}$, $5.10^{-3}$ and $8.10^{-3}$) and three different values of the waiting time $t_w$ ($t_w=10^3$, $10^4$ and $10^5$) at temperature $\mu=\frac{1}{2}$.

\begin{figure}
\centerline{
\epsfxsize = 8.5cm
\epsfbox{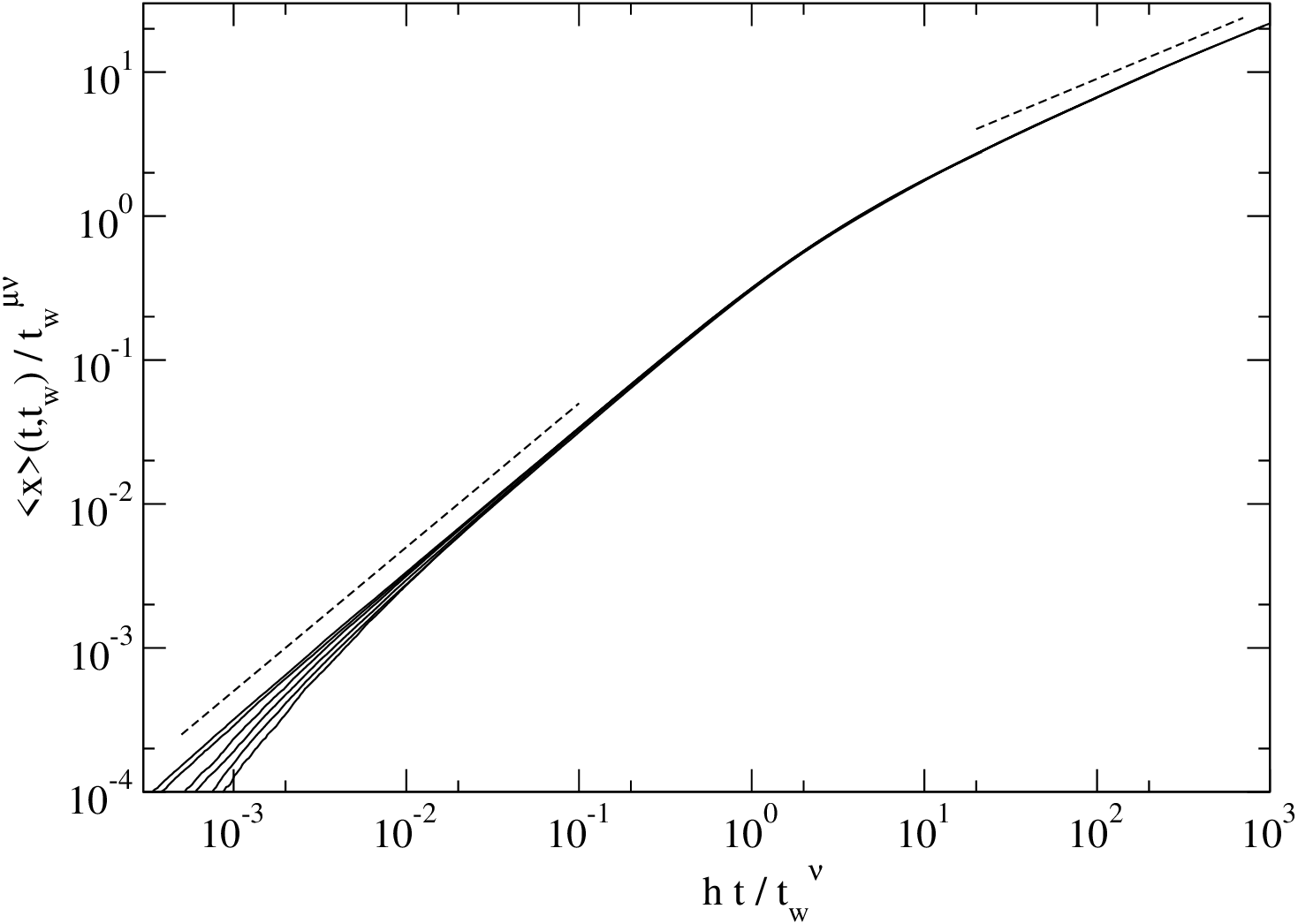}
}
\vskip 0.5 cm
\caption{\sl Rescaled response in the large bias regime ($h \gg h^*(t_w)$), 
using Eq.~(\ref{x-scal-large}), for $h=0.1$, $0.2$, $0.3$, $t_w=10^4$, $10^5$, 
$10^6$, $10^7$ (see text) and $\mu=1/2$, showing a very good collapse of the data,
except for small $t$ where finite time effects become noticeable. The asymptotic
behaviour of the scaling function is also well predicted (dashed lines).
}
\label{fig-large-h}
\end{figure}

In the opposite regime, $h \gg h^*(t_w)$, Eq.~(\ref{x-scal-large}) 
is also well obeyed, as 
shown on Fig.~\ref{fig-large-h} for bias $h=0.1$, $0.2$, $0.3$, 
waiting times $t_w=10^4$, $10^5$, $10^6$, $10^7$ and $\mu=\frac{1}{2}$. 
Note that for clarity, data corresponding to $h=0.2$ and $0.3$ are presented for 
$t_w=10^5$ only. We have checked for several values of $\mu$ that the short time 
and large time behaviour of the scaling functions $f_1(z)$ and $f_2(z)$ given 
by Eqs.~(\ref{f1-behav},\ref{f2-behav}) are also correctly predicted -- 
see Fig.~\ref{fig-small-h} and Fig.~\ref{fig-large-h} for the case $\mu=\frac{1}{2}$.

If one wishes to draw a link with the correlation functions $C(t_w+t,t_w)$ and 
$\Pi(t_w+t,t_w)$ defined above (see also \cite{Bertin02}), it might appear that 
$\la x \ra_h(t,t_w)$ should be associated to 
$C(t_w+t,t_w)$, due to the $t/t_w$ scaling in Eq.~(\ref{x-scal-small}). However,
this is limited to the very small bias case. Interestingly, in the opposite regime 
$h \gg h^*(t_w)$, $\la x \ra_h(t,t_w)$ 
is a function of $t/t^*$ ($t^* \sim t_w^{\nu}/h$), up to a prefactor dependent on 
$t_w$. So at fixed $h$, $\la x \ra_h(t,t_w)$ scales as $t/t_w^{\nu}$, as $\Pi(t_w+t,t_w)$
 does. In fact, this relation is not purely formal, but indeed corresponds to the underlying 
 physics. During the unbiased time interval $[0,t_w]$, the walk typically visits deep traps
 with characteristic time $\sim t_w^{\nu}$.

Once the bias is applied, the evolution is dominated by these deepest traps, before eventually 
reaching a long time regime independent from $t_w$.
If the bias is very small, the walk will 
visit these 
traps a large number of times (of order $t_w^{\mu \nu}$ \cite{Bertin02}), and the aging 
dynamics thus resembles closely that of $C(t_w+t,t_w)$ in the absence of bias. On the contrary,
 if $h$ is large enough, the walk will visit only a few times (of order $1/h$) the deepest traps occupied 
at time $t_w$, so that the aging dynamics is dominated by the time needed to leave the deep traps for the first time, in close analogy with $\Pi(t_w+t,t_w)$. 
Moreover, it is worth mentioning that if one applies the bias from $t=0$ instead of $t_w$, 
and compute the resulting average displacement between times $t_w$ and $t_w+t$, different 
scalings are  obtained. For small fields, the displacement $\langle x \rangle (t,t_w)$ 
behaves as $\langle x \rangle(t,t_w) = ht_w^{2\mu \nu}\, f_1(t/t_w)$ as in the previous case, 
but for larger $h$, one finds a scaling of the form $\langle x \rangle(t,t_w)=t_w^{\mu} f_3(ht/t_w)$. So in this case, a $t/t_w$ scaling also appears, but for a different reason:
 because of the bias, the trapping times reached after a time $t_w$ are now of order $t_w$ 
itself instead of $t_w^{\nu}$, but the walk visits these traps a finite number of times ($\sim 1/h$) after time $t_w$.

\begin{figure}
\centerline{
\epsfxsize = 8.5cm
\epsfbox{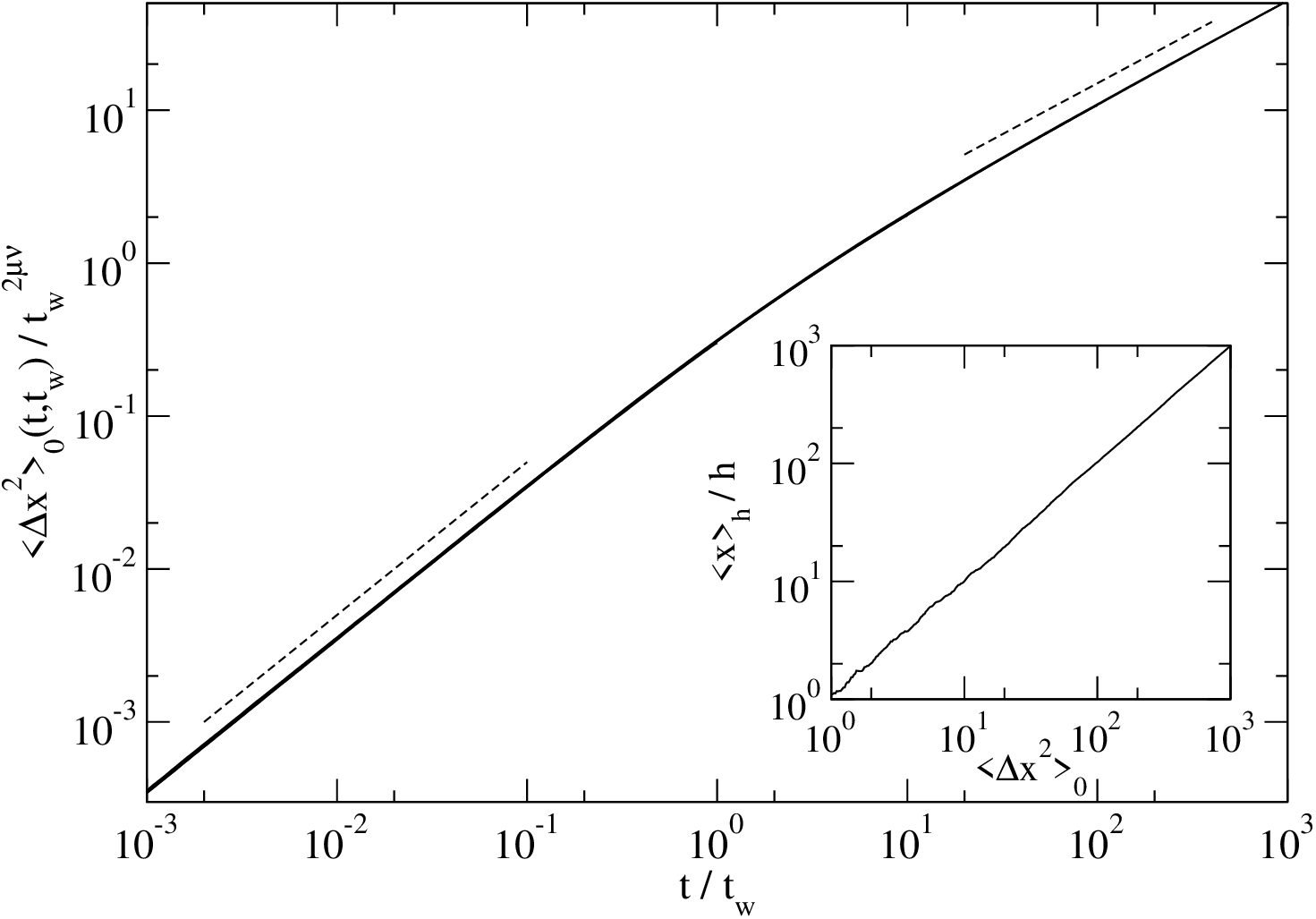}
}
\vskip 0.5 cm
\caption{\sl Rescaled diffusion $\la \Delta x^2 \ra_0(t,t_w)$ 
in the absence of bias, using Eq.~(\ref{x2-sc}), for $t_w=10^3$, 
$10^4$, $10^5$, $10^6$ and $10^7$, and $\mu=1/2$. The different curves 
are indistinguishable to the eye. The dashed lines indicates the slopes 
$1$ and $2/3$ respectively. Note however that for computational reasons, 
time $t$ is limited to $10^7$. Inset: {\sc fdr} plot $\la x \ra_h(t,t_w)/h$ 
versus $\la \Delta x^2 \ra_0(t,t_w)$ for $t_w=10^3$, $h=5.10^{-3}$ and $\mu=1/2$. 
{\sc fdr} is clearly satisfied over the whole range of time.
}
\label{fig-x2}
\end{figure}

Besides the correlation $C$ and $\Pi$ considered hereabove, the natural self-correlation associated to $\la x \ra_h$ is the mean square displacement restricted to the time interval $[t_w, t_w+t]$, in the absence of bias field:
\be
\la \Delta x^2 \ra_0(t,t_w) \equiv \la [x(t_w+t)-x(t_w)]^2 \ra_0
\ee
where the brackets $\langle \cdots \rangle$ denote both a thermal average {\it and} an average over the disorder.
Using the effective trapping time $\overline{\tau}(t_w)$, one can also estimate 
$\la \Delta x^2 \ra_0(t,t_w)$ in the short time, diffusive regime as $\la \Delta x^2 \ra_0(t,t_w) \sim t/\overline{\tau}(t_w)$.
This expression can only be valid if $\la \Delta x^2 \ra_0(t,t_w) \ll \xi(t_w)^2$, 
which yields $t \ll t_w$. In the opposite regime $t \gg t_w$, one recovers the $t_w=0$ 
result $\la \Delta x^2 \ra_0(t,t_w) \sim \xi(t_w+t)^2 \sim \xi(t)^2$. In summary:
\bea
\la \Delta x^2 \ra_0(t,t_w) &\sim& t/t_w^{\nu(1-\mu)} \qquad t \ll t_w \\
\la \Delta x^2 \ra_0(t,t_w) &\sim& t^{2\mu \nu} \qquad \qquad \, t \gg t_w
\eea
This means that $\la \Delta x^2 \ra_0(t,t_w)$ can be written in the scaling form:
\be \label{x2-sc}
\la \Delta x^2 \ra_0(t,t_w) = t_w^{2\mu \nu} g(t/t_w)
\ee
Fig.~\ref{fig-x2} displays the numerical data obtained for 
$\la \Delta x^2 \ra_0(t,t_w)$, for waiting times $t_w$ ranging
from $10^3$ to $10^7$, using Eq.~(\ref{x2-sc}). One can see that the collapse is perfect.

Given the quantities computed above, it is natural to test the Fluctuation-Dissipation (or
Einstein) relation ({\sc fdr}). This is done in the inset of Fig.~\ref{fig-x2}, 
which displays $\la x \ra_h(t,t_w)/h$ versus $\la \Delta x^2 \ra_0 (t,t_w)$ in log-log scale, 
for $h=5.10^{-3}$, $t_w=10^3$ and $\mu=1/2$. This relation appears to be very well 
satisfied over the whole range of time, although the system is strongly out of equilibrium.

We now give a general argument in order to demonstrate the validity of the 
{\sc fdr} for the trap model in the aging regime. It was shown in \cite{PhysRep} that for a given configuration of the disorder and a given initial position $x(t_w)$, the following fluctuation dissipation relation holds, in the limit $h \to 0$:
\be
\langle \Delta x \rangle_h = \langle \Delta x \rangle_0 + h [\langle \Delta x^2 \rangle_0 - \langle \Delta x \rangle_0^2]
\ee
with $\Delta x \equiv x(t_w+t)-x(t_w)$, and $t$ is finite. 
For the purpose of clarity, we distinguish here between average on thermal 
histories after $t_w$ ($\langle \cdots \rangle$) and average over thermal histories 
before $t_w$ -- thus over $x(t_w)$ -- {\it and} over the disorder ($\overline{\cdots}$). Now applying this second type of average to the previous equation, we get:
\be
\overline{\langle x \rangle_h}(t,t_w) = \frac{F}{2kT}\, \overline{\langle \Delta x^2 \rangle_0}(t,t_w)
\ee
where we have used the fact that for the trap model $\la \Delta x \ra_0 = 0$ -- see Eq.~(\ref{posit-current}) --, and $h$ has been replaced by its `physical' expression $F/2kT$. So we conclude that the fluctuation dissipation relation is indeed valid in this out-of-equilibrium and disordered system, with a temperature equal to the bath temperature $T$. In particular, this implies that the scaling functions $f_1(.)$ and $g(.)$ are identical. Note that such an ``aging Einstein relation'' has already been found in the annealed version of the model \cite{Barkai-JCP}.

As a conclusion, we note that the influence of an external bias in disordered system can 
be highly non trivial. In the simple one dimensional trap model discussed here, we already 
find several regimes in the $t,t_w,h$ `cube'. We have extended the above arguments 
to the Sinai model with an external bias (for a review, see \cite{Fisher}). Already for 
$t_w=0$ and small external force $F$, one finds in general four different time regimes for 
the average displacement $\langle x \rangle$. There is in particular a regime where 
$\langle x \rangle$ grows as $\ln^2 t$, but with an $F$ independent prefactor, before the 
asymptotic regime where $\langle x \rangle \sim t^{\alpha F}$ sets in. In other examples, 
such as a walker on a percolation network, the response can even be 
non monotonous with $F$ \cite{PhysRep,BG}.

\end{multicols}

\end{document}